\input harvmac
\input graphicx
%
%
%
%
\ifx\includegraphics\UnDeFiNeD\message{(NO graphicx.tex, FIGURES WILL BE IGNORED)}
\def\figin#1{\vskip2in}
\else\message{(FIGURES WILL BE INCLUDED)}\def\figin#1{#1}
\fi
\def\Fig#1{Fig.~\the\figno\xdef#1{Fig.~\the\figno}\global\advance\figno
 by1}
%
%
%
%
\def\Ifig#1#2#3#4{
\goodbreak\midinsert
\figin{\centerline{
\includegraphics[width=#4truein]{#3}}}
\narrower\narrower\noindent{\footnotefont
{\bf #1:}  #2\par}
\endinsert
}
%
%
\font\ticp=cmcsc10
\def\undertext#1{$\underline{\smash{\hbox{#1}}}$}

\def\hf{{1\over 2}}
\def\calo{{\cal O}}

\def\calh{{\cal H}}
\def\ncol{{\{n\}}}
\def\ncolh{{\{{\hat n}\}}}
\def\ncolp{{\{n'\}}}
\def\cald{{\cal D}}
\def\bhat{{\hat b}}
\def\nhat{{\hat n}}
\def\vhat{{\hat v}}

\def\calA{{\cal A}}

\def\subsubsec#1{\noindent{\undertext { #1}}}
\def\mthsu{\mathsurround=0pt  }
\def\leftrightarrowfill{$\mthsu \mathord\leftarrow\mkern-6mu\cleaders
  \hbox{$\mkern-2mu \mathord- \mkern-2mu$}\hfill
  \mkern-6mu\mathord\rightarrow$}
 \def\overleftrightarrow#1{\vbox{\ialign{##\crcr\leftrightarrowfill\crcr\noalign{\kern-1pt\nointerlineskip}$\hfil\displaystyle{#1}\hfil$\crcr}}}
\overfullrule=0pt
%
%
\lref\Bousrev{
  R.~Bousso,
  ``The holographic principle,''
  Rev.\ Mod.\ Phys.\  {\bf 74}, 825 (2002)
  [arXiv:hep-th/0203101].
}
\lref\tHooholo{
  G.~'t Hooft,
  ``Dimensional reduction in quantum gravity,''
  arXiv:gr-qc/9310026.
  }
\lref\Sussholo{
  L.~Susskind,
  ``The World as a hologram,''
  J.\ Math.\ Phys.\  {\bf 36}, 6377 (1995)
  [arXiv:hep-th/9409089].
}
\lref\tHooftRE{
  G.~'t Hooft,
  ``On The Quantum Structure Of A Black Hole,''
  Nucl.\ Phys.\ B {\bf 256}, 727 (1985).
}
\lref\tHooftFR{
  G.~'t Hooft,
  ``The Black Hole Interpretation Of String Theory,''
  Nucl.\ Phys.\ B {\bf 335}, 138 (1990).
}
\lref\Mald{
  J.~M.~Maldacena,
  ``The large N limit of superconformal field theories and supergravity,''
  Adv.\ Theor.\ Math.\ Phys.\  {\bf 2}, 231 (1998)
  [Int.\ J.\ Theor.\ Phys.\  {\bf 38}, 1113 (1999)]
  [arXiv:hep-th/9711200].
}
\lref\SusskindIF{
  L.~Susskind, L.~Thorlacius and J.~Uglum,
  ``The Stretched horizon and black hole complementarity,''
  Phys.\ Rev.\ D {\bf 48}, 3743 (1993)
  [arXiv:hep-th/9306069].
}
\lref\tHooftRB{
  G.~'t Hooft,
  ``The Black hole horizon as a quantum surface,''
  Phys.\ Scripta {\bf T36}, 247 (1991).
}
\lref\Bous{
  R.~Bousso,
  ``A Covariant Entropy Conjecture,''
  JHEP {\bf 9907}, 004 (1999)
  [arXiv:hep-th/9905177]\semi
  ``Holography in general space-times,''
  JHEP {\bf 9906}, 028 (1999)
  [arXiv:hep-th/9906022].
}
\lref\SGinfo{S.~B.~Giddings,
  ``Quantum mechanics of black holes,''
  arXiv:hep-th/9412138.
}
\lref\BHMR{
  S.~B.~Giddings,
  ``Black holes and massive remnants,''
  Phys.\ Rev.\ D {\bf 46}, 1347 (1992)
  [arXiv:hep-th/9203059].
}
\lref\Astroinfo{
  A.~Strominger,
  ``Les Houches lectures on black holes,''
  arXiv:hep-th/9501071.
}
\lref\CaMa{
  C.~G.~.~Callan and J.~M.~Maldacena,
  ``D-brane Approach to Black Hole Quantum Mechanics,''
  Nucl.\ Phys.\ B {\bf 472}, 591 (1996)
  [arXiv:hep-th/9602043].
}
\lref\Hawkunc{
  S.~W.~Hawking,
  ``Breakdown Of Predictability In Gravitational Collapse,''
  Phys.\ Rev.\ D {\bf 14}, 2460 (1976).
}
\lref\StVa{
  A.~Strominger and C.~Vafa,
  ``Microscopic Origin of the Bekenstein-Hawking Entropy,''
  Phys.\ Lett.\ B {\bf 379}, 99 (1996)
  [arXiv:hep-th/9601029].
}
\lref\GMH{
  S.~B.~Giddings, D.~Marolf and J.~B.~Hartle,
  ``Observables in effective gravity,''
  arXiv:hep-th/0512200.
}
\lref\tHooftFN{
G.~'t Hooft,
 ``Fundamental Aspects Of Quantum Theory Related To The Problem Of Quantizing
Black Holes,''
{\it Prepared for Fundamental Aspects of Quantum Theory: Conference on Foundations of Quantum Mechanics to Celebrate 30 Years of the Aharonov-
Bohm Effect (AB Conference), Columbia, South Carolina, 14-16 Dec 1989}.
}
\lref\tHooftUN{
G.~'t Hooft,
``Horizon operator approach to black hole quantization,''
arXiv:gr-qc/9402037.
}
\lref\VerlindeSG{
E.~Verlinde and H.~Verlinde,
``A Unitary S matrix and 2-D black hole formation and evaporation,''
Nucl.\ Phys.\ B {\bf 406}, 43 (1993)
[arXiv:hep-th/9302022].
}
\lref\SchoutensST{
K.~Schoutens, H.~Verlinde and E.~Verlinde,
``Black hole evaporation and quantum gravity,''
arXiv:hep-th/9401081.
}
\lref\KiemIY{
Y.~Kiem, H.~Verlinde and E.~Verlinde,
``Black hole horizons and complementarity,''
Phys.\ Rev.\ D {\bf 52}, 7053 (1995)
[arXiv:hep-th/9502074].
}
\lref\Hawkrec{
  S.~W.~Hawking,
  ``Information loss in black holes,''
  Phys.\ Rev.\ D {\bf 72}, 084013 (2005)
  [arXiv:hep-th/0507171].
}
\lref\LPSTU{
  D.~A.~Lowe, J.~Polchinski, L.~Susskind, L.~Thorlacius and J.~Uglum,
  ``Black hole complementarity versus locality,''
  Phys.\ Rev.\ D {\bf 52}, 6997 (1995)
  [arXiv:hep-th/9506138].
}
\lref\GiLitwo{
  S.~B.~Giddings and M.~Lippert,
  ``The information paradox and the locality bound,''
  Phys.\ Rev.\ D {\bf 69}, 124019 (2004)
  [arXiv:hep-th/0402073].
}
\lref\Jacob{
  T.~Jacobson,
  ``Introduction to quantum fields in curved spacetime and the Hawking
  effect,''
  arXiv:gr-qc/0308048.
}
\lref\GiLione{
  S.~B.~Giddings and M.~Lippert,
  ``Precursors, black holes, and a locality bound,''
  Phys.\ Rev.\ D {\bf 65}, 024006 (2002)
  [arXiv:hep-th/0103231].
}
\lref\AiSe{
  P.~C.~Aichelburg and R.~U.~Sexl,
  ``On The Gravitational Field Of A Massless Particle,''
  Gen.\ Rel.\ Grav.\  {\bf 2}, 303 (1971).
}
\lref\Penrose{R. Penrose, {\sl unpublished} (1974).}
\lref\EaGi{
  D.~M.~Eardley and S.~B.~Giddings,
  ``Classical black hole production in high-energy collisions,''
  Phys.\ Rev.\ D {\bf 66}, 044011 (2002)
  [arXiv:gr-qc/0201034].
}
\lref\Polchrev{
  J.~Polchinski,
  ``String theory and black hole complementarity,''
  arXiv:hep-th/9507094.
}
\lref\BPS{
  T.~Banks, L.~Susskind and M.~E.~Peskin,
  ``Difficulties For The Evolution Of Pure States Into Mixed States,''
  Nucl.\ Phys.\ B {\bf 244}, 125 (1984).
}
\lref\SGloc{S.B. Giddings, ``Locality in quantum gravity and string theory," arXiv:hep-th/0604072.}
\lref\BaFi{T.~Banks and W.~Fischler,
  ``A model for high energy scattering in quantum gravity,''
  arXiv:hep-th/9906038.
}
\lref\GiTh{
  S.~B.~Giddings and S.~D.~Thomas,
  ``High energy colliders as black hole factories: The end of short  distance
  physics,''
  Phys.\ Rev.\ D {\bf 65}, 056010 (2002)
  [arXiv:hep-ph/0106219].
}
\lref\DiLa{
  S.~Dimopoulos and G.~Landsberg,
  ``Black holes at the LHC,''
  Phys.\ Rev.\ Lett.\  {\bf 87}, 161602 (2001)
  [arXiv:hep-ph/0106295].
}
\lref\CGHS{
  C.~G.~Callan, S.~B.~Giddings, J.~A.~Harvey and A.~Strominger,
  ``Evanescent black holes,''
  Phys.\ Rev.\ D {\bf 45}, 1005 (1992)
  [arXiv:hep-th/9111056].
}
\lref\GiNe{
  S.~B.~Giddings and W.~M.~Nelson,
  ``Quantum emission from two-dimensional black holes,''
  Phys.\ Rev.\ D {\bf 46}, 2486 (1992)
  [arXiv:hep-th/9204072].
}
\lref\Hawkevap{
  S.~W.~Hawking,
  ``Particle Creation By Black Holes,''
  Commun.\ Math.\ Phys.\  {\bf 43}, 199 (1975)
  [Erratum-ibid.\  {\bf 46}, 206 (1976)].
}
\lref\Wald{
  R.~M.~Wald,
  ``On Particle Creation By Black Holes,''
  Commun.\ Math.\ Phys.\  {\bf 45}, 9 (1975).
}
\lref\Waldunpub{R.~M.~Wald, unpublished.}
\lref\HoMa{
  G.~T.~Horowitz and J.~Maldacena,
  ``The black hole final state,''
  JHEP {\bf 0402}, 008 (2004)
  [arXiv:hep-th/0310281].
}
\lref\ChFu{
  S.~M.~Christensen and S.~A.~Fulling,
  ``Trace Anomalies And The Hawking Effect,''
  Phys.\ Rev.\ D {\bf 15}, 2088 (1977).
}
\lref\GiddingsXY{
  S.~B.~Giddings and V.~S.~Rychkov,
  ``Black holes from colliding wavepackets,''
  Phys.\ Rev.\ D {\bf 70}, 104026 (2004)
  [arXiv:hep-th/0409131].
}
\lref\MagueijoAM{
  J.~Magueijo and L.~Smolin,
  ``Generalized Lorentz invariance with an invariant energy scale,''
  Phys.\ Rev.\ D {\bf 67}, 044017 (2003)
  [arXiv:gr-qc/0207085].
}
\lref\Amel{
  G.~Amelino-Camelia, C.~Lammerzahl, A.~Macias and H.~Muller,
  ``The search for quantum gravity signals,''
  AIP Conf.\ Proc.\  {\bf 758}, 30 (2005)
  [arXiv:gr-qc/0501053].
}
\lref\Amelino{
  G.~Amelino-Camelia, M.~Arzano, Y.~Ling and G.~Mandanici,
  ``Black-hole thermodynamics with modified dispersion relations and
  generalized uncertainty principles,''
  arXiv:gr-qc/0506110.
}
\lref\Kowal{
  J.~Kowalski-Glikman,
  ``Doubly special relativity: Facts and prospects,''
  arXiv:gr-qc/0603022.
}
\lref\FHKS{
  L.~Fidkowski, V.~Hubeny, M.~Kleban and S.~Shenker,
  ``The black hole singularity in AdS/CFT,''
  JHEP {\bf 0402}, 014 (2004)
  [arXiv:hep-th/0306170].
}
\lref\HossenfelderCW{
  S.~Hossenfelder,
  ``Interpretation of quantum field theories with a minimal length scale,''
  Phys.\ Rev.\ D {\bf 73}, 105013 (2006)
  [arXiv:hep-th/0603032].
}
\lref\Bohr{N.~Bohr, {\sl Collected works}, Vol. 2, p. 136 (North Holland, Amsterdam).}
\Title{\vbox{\baselineskip12pt
\hbox{hep-th/0605196}
}}
{\vbox{\centerline{Black hole information, unitarity, and nonlocality}
}}
\centerline{{\ticp Steven B. Giddings}\footnote{$^\star$}{Email address: giddings@physics.ucsb.edu} 
}
\centerline{ \sl Department of Physics}
\centerline{\sl University of California}
\centerline{\sl Santa Barbara, CA 93106-9530}
\bigskip
\centerline{\bf Abstract}
The black hole information paradox apparently indicates the need for a fundamentally new ingredient in physics.  The leading contender is  nonlocality.  Possible mechanisms for the nonlocality needed to restore unitarity to black hole evolution are investigated.  Suggestions that such dynamics arises from  ultra-planckian modes in Hawking's derivation are investigated and found not to be relevant, in a picture using smooth slices spanning the exterior and interior of the horizon.  However, no simultaneous description of modes that have fallen into the black hole and outgoing Hawking modes can be given without appearance of a large kinematic invariant, or other dependence on ultra-planckian physics; a reliable argument for information loss thus has not been constructed.  This suggests that strong gravitational dynamics is  important.  Such dynamics has been argued to be fundamentally nonlocal in  extreme situations, such as those required to investigate the fate of information.

\Date{}

\newsec{Introduction}

Hawking's argument that black holes destroy information\refs{\Hawkunc} and thus violate quantum mechanics initiated a crisis  in physics.  In short, in the wake of this analysis, there is no apparent way to reconcile 
the basic principles of locality and energy conservation within known extensions of quantum physics.  A paradox results.

One should consider the possibility that this crisis is of equal importance  to the ultraviolet crises in classical physics that were only resolved by the advent of quantum mechanics.  In particular, one might compare the black hole paradox to that of the classical instability of matter, which was only avoided by introduction of a fundamentally new ingredient,  quantization.

If this analogy holds, we should seek this fundamentally new ingredient, beyond the principles of local quantum field theory and semiclassical gravity, in order to resolve the present paradox.  This fundamental new ingredient may represent the beginning of a transition as conceptually profound as that from classical physics to quantum physics.

A leading possibility is that this new ingredient is some sort of nonlocality.   This idea was advocated by the present author in \refs{\BHMR}, in 't Hooft's seminal work\refs{\tHooholo},  and has been
championed and greatly elaborated by Susskind\Sussholo.  In particular, it has become widely believed that the Bekenstein-Hawking entropy, which grows with the surface area, is an accurate measure of the number of degrees of freedom inside a black hole, in contrast to volume-growth predictions of local quantum field theory, and this  ``holographic principle" has been refined and extended in conceptual work of Susskind, Bousso, and others\refs{\Bous-\Bousrev}.  Moreover, the related proposal of  ``black hole complementarity"\refs{\tHooftRE\tHooftFR\tHooftRB-\SusskindIF}, which states that there is no way for observers inside and outside a black hole to compare information content and thus reveal a contradiction if information escapes, and thus that their pictures may be complementary, has been proposed as part of the rationale for such a holographic viewpoint.  Elegant calculations of black hole entropy from string theory\refs{\StVa,\CaMa}, as well as indications from the AdS/CFT correspondence\refs{\Mald} have also buttressed this viewpoint.

Despite these developments, and a growing consensus that locality is not fundamental, one needs  to understand   where precisely Hawking's calculation fails, in order to actually resolve the paradox.  One needs to better characterize nonlocal physics, ideally in terms of some mechanism, together with a description of where this mechanism is operative, and why it is relevant to black hole physics.  

One possible origin of nonlocality in string theory is the extended nature of strings.  However, despite suggestions that such nonlocality could resolve the paradox (see {\it e.g.} \LPSTU), studies of high-energy scattering in string theory show no evidence of nonlocality on scales that would correspond to production of the requisite long strings\SGloc.\foot{The possibility of some nonlocal behavior on scales intermediate between those of stretched strings and of gravity has not been eliminated, and such nonlocal phenomena could possibly be relevant\SGloc; however, gravitational physics seems to retain a central role.}  

On the other hand, there is a more generic possibility:  the requisite nonlocality is an intrinsic feature of strong gravitational dynamics.  This viewpoint\refs{\GiLione\GiLitwo-\GMH,\SGloc} emerges from apparent gravitational limits in Gedanken experiments where one tries to localize very high-energy particles.  In essence, the  proposal (which might be termed a {\it nonlocality principle}) is that gravitational dynamics is fundamentally quantum-mechanical, but the dynamics that unitarizes gravity in strongly-coupled regimes is nonlocal, or perhaps more precisely, has no intrinsic notion of locality.  If this is the case, this dynamics presumably cannot be derived from quantum field theory and general relativity (and possibly not from string theory), just as quantum mechanics can't be derived from classical mechanics.  

The physics that we see experimentally is so far manifestly local, and moreover locality is an important ingredient for consistency of quantum field theory.  Thus any such nonlocality should be constrained to only appear in extreme circumstances beyond the realm of previous investigations.  In particular, one criterion for its appearance, in terms of {\it locality bounds}\refs{\GiLione,\GiLitwo,\SGloc}, clearly indicates that it would only be manifest in situations where one considers superplanckian energies.

If such a nonlocality principle plays a role in black hole physics, we should seek an explanation for how the requisite extreme circumstances arise in the black hole context.  There has long been a sense that the issue may lie in the peculiar dynamics of Hawking radiation\Hawkevap, which refers to modes close to the black hole that have ultrahigh energies in the frame of an infalling observer\refs{\tHooftRE,\tHooftFN,\tHooftFR,\tHooftUN\VerlindeSG-\SchoutensST}.  There is a counterargument to this, known as the ``nice slice argument,"\foot{This type of argument, which has had a long history in the discussion of the subject, was particularly well-codified by \refs{\Waldunpub,\LPSTU,\Polchrev,\Jacob}.}
from which viewpoint there is no physical role for ultraplanckian modes.  This paper examines this question more carefully, in particular giving a calculation reconciling descriptions of outside observers with evolution on nice slices;   no role is found for  such ultraplanckian modes. 

However, there {\it do} appear to be ultraplanckian invariants, in comparing Hawking modes -- once they become ``real," at energies  comparable to the Hawking temperature -- to modes of the infalling matter in the black hole interior.  An argument for information loss must do precisely this, namely must describe the state  of infallen information and the outgoing radiation simultaneously on a spacelike slice; the standard argument for a mixed density matrix then follows from tracing over the part of the state inside  the black hole.  This is a situation where it is so far not possible to justify use of semiclassical local physics.  Assumption of local physics moreover leads to the paradox.  Thus this is plausibly a situation where the proposed nonlocal physics enters the calculation of properties of Hawking radiation, restores purity, and removes the paradox.  (A different viewpoint on the resolution to the paradox is presented in \refs{\Hawkrec}.)

In outline, the next section  reviews the basic argument for information loss, and why this results in a paradox.  Section three then provides a more detailed description of Hawking evaporation, explaining the reasoning behind the semiclassical approximation, and sketching the calculation of the density matrix.  The apparent role of ultraplanckian modes in the Hawking radiation is investigated; it is found that while such modes appear to contribute to the stress tensor, they in fact have no physical effect on, for example, infalling particles.  This is explained in terms of the failure of such modes and counterpart modes inside the horizon to separate from the horizon and from each other;
apparently Hawking modes only become ``real" and interacting once they reach energies characterized by the Hawking temperature.  Section four then turns to the question of a possible role for nonlocal dynamics, first explaining generalities of how it could emerge in gravitational physics.  The kinematics of infalling particles and outgoing Hawking radiation is then investigated, and found to produce large invariants that suggest a role for such nonlocal dynamics.  Possible relations of such a picture to holography, complementarity, and the final state proposal of \HoMa\ are also discussed.  The paper then closes with a summary, together with some more perspective on the paradox, the role of nonlocal physics, and the possible analogy to the transition from classical physics to quantum physics.

\newsec{Tomography, information loss, and paradox}

Let us recall the basics of the arguments that lead to a paradox. This will set the stage for investigating how it might be evaded.

The central question first raised by Hawking\refs{\Hawkunc} is whether pure states evolve to mixed in the context of black hole formation and evaporation.  This appears to be a meaningful question, which could in principle be tested by some future experimentalist; in the most optimistic of worlds black hole production could begin with the LHC\refs{\BaFi\GiTh-\DiLa}. The test would be through a process of  ``quantum tomography."  One could imagine preparing a sequence of identical initial pure quantum states, {\it e.g.} of colliding particles, which produce black holes.  According to Hawking, these evaporate into radiation  consisting of outgoing particles of various species in an apparently thermal distribution.  

Our ambitious experimentalist might try to measure probabilities for various configurations of this outgoing radiation.  Specifically, she may have apparatus that detects particles in various modes, {\it e.g.} wavepackets with approximately definite momentum and position.  The corresponding out states can be labeled in occupation number basis as $|\{n_i\}\rangle$ where $i$ denotes the mode in question.  The information content is determined by carefully determining the elements of the density matrix,
\eqn\denselem{\rho_{\ncol\ncolp} = \langle \ncol|\rho | \ncolp\rangle\ ,}
in this series of experiments.
The quantity that gives a precise measure of this information is the entropy, 
\eqn\entdef{S=-Tr(\rho\log\rho)\ .}
If our experimentalist determines the elements of the density matrix by suitable projections onto outgoing states, in repeated  experiments, she would conclude  the state is pure if $S=0$, or mixed if $S\neq0$.  

Note that a definite conclusion requires measurement of all the elements of the density matrix. For example, values of individual matrix elements can make the difference between pure and mixed states, as comparison of the mixed state 
\eqn\exone{\rho_1 = \hf\left(|0\rangle\langle0| + |1\rangle\langle1|\right)}
and pure state
\eqn\extwo{\rho_2 = \hf \left( |0\rangle + |1\rangle\right) \left( \langle 0| + \langle1|\right)\ ,}
illustrate.

Hawking prompted our  conceptual crises in physics by arguing that the density matrix for such black hole decay is mixed\refs{\Hawkunc}.  The essence of his argument is to study the evolution of an initial pure state $|\Psi_i\rangle$ into a state on a spatial slice that spans both interior and exterior of the black hole, and thus intersects both the modes that have carried information into the black hole, and the outgoing modes of Hawking radiation.  One assumes, in accord with local field theory, that the Hilbert space on such a slice can be decomposed into a tensor product
\eqn\hilbprod{\calh = \calh_{inside}\otimes \calh_{outside}\ .}
Denoting a general basis of inside and outside states by $|\alpha\rangle$ and $|a\rangle$, 
Hawking's computation yields a definite state of the form
\eqn\Hawkstate{|\Psi_i\rangle = \sum_{a\alpha} \Psi_{i,a\alpha} |a\rangle |\alpha\rangle\ }
on the spatial slice.
Measurements made outside the black hole are summarized by the density matrix found by tracing over the inside states $|\alpha\rangle$.  As we will review, Hawking's argument predicts that this density matrix is approximately thermal, and has a correspondingly large entropy.  But these details are not needed to see the essence of the problem -- locality indicates that there will be such inside states, with degrees of freedom independent of states outside the horizon, and that they will be excited by infalling matter.  Thus the outside density matrix of an initially pure state must have non-vanishing entropy.

This raises a serious problem.  If quantum information can be lost in the formation and evaporation of a real black hole, on general quantum grounds there is  nothing that would forbid its taking place in virtual processes.  Production of virtual black holes should in particular be unsuppressed at the Planck scale.  So information loss would take place all the time.  Yet information and energy are inextricably linked -- relaying a bit of information in a time $\Delta t$ requires an energy $1/\Delta t$ -- and so such information loss would lead to catastrophic breakdown of energy conservation.  This is discussed in more detail in \refs{\BPS}, which describes models of evolution with information loss.

If the Hawking radiation does not contain information, the only alternative that could save unitary evolution is for the information to be preserved in some form of black hole remnant that is left behind when the semiclassical approximation apparently breaks down at black hole mass $M\sim M_{pl}$.  This implies a new type of object, with mass $M\sim M_{pl}$, and with an {\it infinite} number of internal states to encode the infinite varieties of information that could be fed to a black hole.  The connection between information and energy ensures that such remnants would be very long-lived, given the small energy that would be available to carry away the large remnant information in its decay.
This, too, is a disaster -- due to their infinite degeneracy, such remnants would be infinitely produced in generic physical processes.

Thus very general principles of local quantum physics and general relativity lead to a paradox.

\newsec{Dynamics of black hole decay}

\subsec{Semiclassical approximation}

A complete calculation of the matrix elements $\langle a\alpha |\Psi_i\rangle$ of \Hawkstate\ requires a full theory of quantum gravity.  In the absence of such a theory in which we can perform this calculation,\foot{If string theory is such a theory, it's current status doesn't permit such calculations.} note that outside the regime where planckian effects are relevant, the calculation can be treated as a functional integral over the metric and relevant matter degrees of freedom.  A central point in the argument for information loss\Hawkunc\  is that it can be reliably made in a semiclassical approximation for the metric.  Thus, even if the functional integral over metrics is only an approximation to more fundamental dynamics, it should serve as a reliable indicator of its own validity; it should be valid in the semiclassical regime, and moreover its breakdown points to where more fundamental dynamics is required.

Thus we represent the amplitude as
\eqn\fctlint{\langle a\alpha|\Psi_i\rangle = \int_{\Psi_i[g,\phi]}^{\Psi_{a\alpha}[g,\phi] }
 \cald g \cald \phi e^{i S[g]+iS[g,\phi]}\ .}
Here we must bear in mind that the  initial and final states $\Psi_i$, $\Psi_{a\alpha}$ give data for both matter and metric configurations, 
which are of course related by constraints.  The action has been decomposed into the purely metric part, $S[g]$, and the (metric-dependent) matter contribution $S[g,\phi]$.

A basic point of the argument for information loss\Hawkunc\ is that in a more-fundamental calculation like this, the backreaction from Hawking radiation on a large black hole is small, and can be treated as an appropriate average; as a result, the metric can be replaced by a semiclassical, slowly evolving metric.  Indeed, this result follows from the expression \fctlint\ if, as expected, for a typical state $|a\alpha\rangle$ of the Hawking radiation, we can approximate
\eqn\approxgrav{\int_{\psi_i}^{\psi_{a \alpha}}  \cald \phi e^{iS[g,\phi]} \approx e^{iS_{HR}[g]}\ ,}
where $S_{HR}[g]$ is an effective action (which depends on the initial state) summarizing the effects of the Hawking radiation.  Lowercase $\psi$ denotes the matter part of the state $\Psi$.  For a suitable average over states $|a\alpha\rangle$, this step can be made quite explicit, for example in the two-dimensional CGHS model\refs{\CGHS}.  A test of the near independence on the specific matter state will be described shortly.

For a large mass hole, the initial state specified through $\Psi_i$ corresponds to a nearly classical metric $g_i$
 and the backreaction from $S_{HR}$ is weak, so the functional integral over $g$ should be well approximated by a saddlepoint expression, about the classical solution $\bar g$ satisfying the equation of motion
\eqn\classmet{{\delta\over \delta g_{\mu\nu}} \left( S[g]+ S_{HR}[g]\right)_{\big\vert{\bar g}} = 0\ .}
The amplitude \fctlint\ then takes the form
\eqn\approxint{\langle a\alpha|\Psi_i\rangle \approx A e^{iS[\bar g]} \int _{\psi_i}^{\psi_{a \alpha}}  \cald \phi e^{iS[\bar g,\phi]}\ }
where $A$ includes the one-loop determinant from the integral over $g$.

The matter functional integral in \approxint, in the background classical metric $\bar g$, has been approximately computed in various ways, beginning with \Hawkevap.  Suppose, for example, we work with a non-interacting scalar field $\phi$.  Decompose this field in terms of modes appropriate to representing the initial state, or alternately the combined final state inside and outside the black hole:
\eqn\modedecomp{\eqalign{\phi &= \sum_i a_i u_i + a^\dagger_i u^*_i \quad {\rm (in)}\cr
&= \sum_i b_i v_i + b^\dagger_i v^*_i + \sum_\iota \bhat_\iota \vhat_\iota + \bhat^\dagger_\iota \vhat^*_\iota\quad {\rm (out + internal)}\ .}}
Here the $u_i$, $v_i$, $\vhat_\iota$ are bases of modes, chosen to be positive frequency in some appropriate convention, for the respective regions ``in," ``out," and ``inside," and the $a_i$, $b_i$, and $\bhat_\iota$ are corresponding annihilation operators.  In this case, it is quite natural to decompose the final states in an occupation number basis, $|\{n_i\},\{\nhat_\iota\}\rangle$.  In this basis the wavefunction takes the form
\eqn\psiapprox{|\Psi_i\rangle\approx \sum_{ \{n\},\{\nhat\} }|\{n\},\{\nhat\}\rangle \langle \{n\},\{\nhat\}|\Psi_i\rangle\ }
where \approxint\ computes the individual amplitudes.  More discussion will be given of the detailed form of these amplitudes, for which quite explicit expressions can be given, again in appropriate approximations.  But the essential point is that this representation of the state shows a high degree of entanglement between the internal and external degrees of freedom, so the density matrix
\eqn\densocc{\rho_{ \{n\} \{n'\}} = \sum_{\{\nhat\}} \langle \{n\},\{\nhat\}|\Psi_i\rangle \langle \Psi_i |\{n'\},\{\nhat\}\rangle}
has large non-vanishing entropy.  In fact, one finds that with the approximate expressions for the amplitudes, the entropy agrees with the expected value for a thermal state.  Since there is no apparent means to recover the information lost to the interior, Hawking\Hawkunc\ argued that the entropy of \densocc\ represents information that has been fundamentally lost in the black hole formation and evaporation, and quantum mechanical evolution breaks down.

A more careful characterization of the Hawking radiation, and check of the justification of the semiclassical approximation, arises from derivation of the matrix elements in \psiapprox, to which we proceed next.  

\subsec{Aspects of Schwarzschild geometry}

Begin by considering a black hole of mass $M$ that forms from collapsing matter; we will work with the non-rotating, spherically symmetrical case.  We can write the metric in the form
\eqn\confmet{ds^2 = f(r_*,t)(-dt^2 + dr_*^2) + r^2(r_*,t) d \Omega^2\ .}
After the black hole forms, the metric is classically Schwarzschild.  We identify $t$ as the asymptotic Schwarzschild time, and in this future region $r_*$ as the usual tortoise coordinate.  Corresponding retarded and advanced coordinates can be defined by
\eqn\retadv{u=t-r_*\quad ,\quad v=t+r_*\ .}

Description of infalling observers is accomplished by introducing Kruskal coordinates,
\eqn\kruskdef{ U= - 4Me^{-u/4M-1/2}\quad ,\quad V=4Me^{v/4M-1/2}\ ,}
in terms of which the Schwarzschild metric becomes
\eqn\kruskmet{ds^2= -{2M\over r} e^{1-(r/2M)}dUdV + r^2 d\Omega_2^2\ ,}
where $d\Omega_2^2$ is the line element on $S^2$.  Use of these coordinates explicitly shows that the vicinity of the horizon, $|r-2M|\roughly< M$, is equivalent to flat space, with $t$ mapping to Rindler time.  We refer to this near-horizon region as the {\it Rindler region}. 

\subsec{Characteristics of Hawking radiation}

Outside modes of a massless scalar field $\phi$ can be written
\eqn\modefour{v_{\omega lm} \propto e^{-i\omega t} R_{\omega l}(r_*) Y_{lm}(\theta,\phi)}
in the future region.  The $Y_{lm}$ are the usual spherical harmonics.  Different bases arise from different choices of boundary conditions for $R_{\omega l}$.  The modes useful for computing the Hawking flux have boundary conditions\refs{\Wald,\Hawkunc}
\eqn\upmodes{\eqalign{rR_{\omega l} &\rightarrow t_{\omega l} e^{i\omega r_*} \quad ,\quad\quad\quad\quad\quad r_*\rightarrow\infty\ \ ;\cr &\rightarrow e^{i\omega r_*} + r_{\omega l} e^{-i\omega r_*}\quad ,\quad   r_*\rightarrow-\infty\ ({\rm horizon})\ ,}}
where $t_{\omega l}$ and $r_{\omega l}$ are transmission and reflection coefficients in the effective potential for $R_{\omega l}$.    

While for some purposes discussion is simplified by working in such a plane-wave basis, for many purposes it is better to work with a wavepacket basis.  This allows one to give an approximately local description of the particle states created.  This is important, for example, as the black hole actually has a finite lifetime, which has the important effect of cutting off the total energy and entropy of the Hawking radiation.
There are many ways to construct acceptable wavepackets from positive frequency modes, they may be gaussians, square wavepackets
\eqn\sqw{v_{kn}={1\over\sqrt\epsilon} \int_{k\epsilon}^{(k+1)\epsilon} d\omega e^{2\pi i \omega n/\epsilon }v_\omega}
as described in  \refs{\Hawkevap,\GiNe}, or other functions localized in position and frequency.  
Let the wavepacket modes be denoted $v_K$, where the index $K$ includes the angular momenta $l,m$.  

The derivation of the outgoing Hawking state can be carried out either by tracing such wavepackets back through the collapsing body that formed the black hole, or alternatively by using the fully extended Schwarzschild geometry.  In either case, if a wavepacket $v_K$ corresponding to an outgoing Hawking particle is traced back along the horizon to either the origin or the past horizon (depending on which geometry we use), it is dominated by the outgoing piece in \upmodes.  In the case of collapse, we for simplicity assume that the scalar field starts in its vacuum $|0\rangle_{in}$; for example the collapsing matter could be of a different kind.  In the fully extended geometry, the equivalent situation can be described using the Kruskal coordinates \kruskdef\ 
by requiring positive frequency with respect to $U$ on the past horizon.  There, the radial wavefunction
\upmodes\ asymptotes to
\eqn\outasy{rR_{\omega l}e^{-i\omega t}\rightarrow  e^{-i\omega u} = \left(-{U\sqrt e\over 4M}\right)^{4iM\omega}\ .}
 One can likewise
define inside modes ${\hat v}^*_{\omega l}$ with asymptotic behavior 
\eqn\insideasy{ r{\hat R^*}_{\omega l}e^{i\omega t}\rightarrow  \left({U\sqrt e\over 4M}\right)^{4iM\omega}}
at the horizon.  Then positive frequency modes in $U$ can be found by analytic continuation.  The expression \outasy\ has a branch point at the horizon, $U=0$.  Analyticity in the lower-half $U$ plane corresponds to positive frequency in $U$, so continue around this point using the contour $U=-\epsilon e^{i\theta}$, with $0\leq\theta\leq\pi$.  Restricting to the real line results in the positive frequency (in $U$) modes
\eqn\vfd{\eqalign{v^1_{\omega lm} &= \left[v_{\omega lm}(U) \theta(-U) + \gamma_\omega {\hat v}^*_{\omega lm}(U)\theta(U)\right]/\sqrt{1-\gamma_\omega^2}\cr v^2_{\omega lm} &= \left[{\hat v}_{\omega lm}(U) \theta(U) + \gamma_\omega v^*_{\omega lm}(U)\theta(-U)\right]/\sqrt{1-\gamma_\omega^2}\ ,
}}
where
\eqn\gammafd{\gamma_\omega = e^{-4\pi M\omega}\ .}
Here we only describe the inside modes in the region where the metric is Schwarzschild.  Their past origin depends on the past extension of the metric; if this were the extended Schwarzshchild solution, they would correspond to modes originating behind the horizon, in the other asymptotic region, but in a collapse situation they arise from certain incoming modes from $\cal I^-$ that bounce off $r=0$ behind the horizon.

In an expansion \modedecomp\ in terms of the modes $v^1_{\omega lm}$,  $v^2_{\omega lm}$, the corresponding annihilation operators must annihilate the in-state $|0\rangle_{in}$; using \vfd\ these can be written in terms of the ``out" and ``inside" operators as
\eqn\bogolxm{\eqalign{a^1_{\omega lm}&=(1-\gamma_{\omega }^2)^{-1/2} \left(b_{\omega lm} - \gamma_\omega \bhat_{\omega lm}^\dagger\right)\cr a^2_{\omega lm}&=(1-\gamma_\omega^2)^{-1/2} \left(\bhat_{\omega lm} - \gamma_\omega b_{\omega lm}^\dagger\right)\ .}
}

One easily finds the state annihilated by these, in terms of the occupation number basis $\{n_{\omega lm}\}$ and $\{\nhat_{\omega lm}\}$ for 
the ``out" and ``inside" Hilbert spaces, respectively:
\eqn\prodH{|0\rangle_{in} = C \sum_{\ncol,\ncolh} e^{-{4\pi M}\sum_{lm}\int d\omega \omega n_{\omega lm}} \delta_{\ncol\ncolh}|\{n\}\rangle |\ncolh\rangle\ ;}
notice that in the basis we have defined, occupation numbers of inside and outside modes are always equal.  $C$ is a normalization constant.
The corresponding density matrix 
\eqn\densmat{\rho_{\ncol\ncolp}=C^2 \delta_{\ncol\ncolp}e^{-{8\pi M}\sum_{lm}\int d\omega \omega n_{\omega lm}}\ }
is thermal with Hawking temperature given by
\eqn\Thawk{T_H^{-1}=\beta_H = {8\pi M}\ .}

The expressions \prodH\ and \densmat\ are actually not quite correct, as the states on the right hand side have both infinite energy and entropy.   A more precise expression comes from working with the wavepacket states, and accounting for the backreaction, which causes the black hole to evaporate in finite time.  The expected modifications to \prodH\ are then A) the occupation numbers summed over should be those of the wavepacket modes, $\{n_K\}$, $\{{\hat n}_K\}$, and B) 
description of these modes should take into account the gradual shrinkage and eventual disappearance of the black hole through evaporation; this yields an effectively time-dependent temperature, and ultimate cutoff on the Hawking process.  The resulting density matrix is expected to have finite energy and entropy.

\subsec{Ultra-high energies and nice slices}

Hawking's original derivation, whose streamlined version has been given above, apparently makes reference to ultra-high energy modes.  Specifically, the wavepackets of a given outgoing state $|\{n_K\}\rangle$, when traced back to their origin near the horizon, are highly blueshifted as seen by an infalling observer.  The apparent dependence of the calculation on properties of such modes raises suspicion about its validity.  Indeed, work of 't Hooft\refs{\tHooftRE,\tHooftFN,\tHooftFR,\tHooftUN} and other subsequent work\refs{\VerlindeSG,\SchoutensST}, argued that such blueshifts would play a critical role in a breakdown of Hawking's argument for information loss.  

A good way to explore this issue is to investigate the stress tensor arising from these modes near the horizon.  In particular, in justifying the semiclassical approximation, near independence of the backreaction on the outgoing state was used, \approxgrav.  But, if the large blueshift is important, one expects small differences in the final outgoing state to magnify as it is traced back to near the horizon.  Specifically, consider the matrix element of the quantum stress tensor,
\eqn\qstress{  {\delta\over \delta g^{\mu\nu} }\int_{\psi_i}^{\psi_{a \alpha}}  \cald \phi e^{iS[g,\phi]} =\langle a\alpha |T_{\mu \nu} |\psi_i\rangle\ ;}
this is a more accurate description of the source for the metric than the average stress tensor in \classmet, which can be written  
\eqn\tav{{T}^0_{\mu\nu} = \langle \psi_i |T_{\mu\nu} | \psi_i\rangle\ .}
The difference between the two, in a state of definite occupation numbers,
\eqn\tdiffer{\langle \{n\},\{\hat n\}|\Delta T_{\mu\nu} |0\rangle_{in} = \langle \{n\},\{\hat n\}| (T_{\mu\nu} -T^0_{\mu\nu}) |0\rangle_{in}\ ,}
can be evaluated from the stress tensor
\eqn\fdstress{T_{\mu\nu} = \hf \left[ \partial_\mu \phi \partial_\nu \phi - \hf g_{\mu\nu} (\partial \phi)^2\right]\ }
and the mode expansion \modedecomp.  An advantage to working with this difference is that normal-ordering dependence cancels in the difference.  While difficult to explicitly evaluate in four dimensions, $T^0_{\mu\nu}$ can be explicitly evaluated in two-dimensional models\refs{\ChFu,\CGHS,\GiNe}.
The ``outside" stress tensor, normal ordered with respect to the outgoing modes, takes the form
\eqn\fdstexp{:T_{\mu\nu}: = \sum_{K,L} \left[\hf t_{\mu\nu}(v_K,v_L)b_K b_L + \hf t_{\mu\nu}(v^*_K,v^*_L)b_K^\dagger b_L^\dagger + t_{\mu\nu}(v^*_K,v_L) b^\dagger_K b_L\right]}
where
\eqn\tmndef{ t_{\mu\nu}(f,g) =  \partial_{(\mu} f \partial_{\nu)} g - \hf g_{\mu\nu} \partial f\cdot \partial g\  .}
A similar expression, in terms of the operators $\bhat_K$, holds inside the black hole.

The difference \tdiffer\ can be evaluated using \fdstexp; the first two terms, and contributions to the third with $K\neq L$, approximately cancel due to large phases and/or small wavepacket overlaps, resulting in
\eqn\fdT{\langle \{n\},\{\hat n\}|\Delta T_{\mu\nu} |0\rangle_{in} = Ce^{-\beta_H E(\{n\})/2} \left[ \sum_K t_{\mu\nu}(v^*_K,v_K)n_K - T^0_{\mu\nu}\right]\delta_{\ncol\ncolh}\ .}
Appropriate coordinates for an infalling observer in the vicinity of the horizon are for example the Kruskal coordinates $U,V$ of eq.~\kruskdef, since, as we've seen,  they approximate Minkowski coordinates.  A typical mode $v_K$ in \fdT\ has frequency $\sim T_H\sim 1/M$ in the Schwarzschild coordinates $u,v$.  For late retarded times $u$, this gets converted to a frequency that is exponentially large in the Kruskal time seen by the infalling observer. Thus, in \fdT, deviations from the average state are indeed greatly magnified, suggesting a breakdown of the semiclassical analysis.

\subsubsec{Interactions with infalling matter}

This assertion can be tested by examining gravitational interactions of infalling matter with this ``thermal atmosphere" of the black hole. Specifically, consider the amplitude of an infalling $\phi$ particle of momentum $p$ to interact with the atmosphere, resulting in an infalling particle of momentum $p'$ and an outgoing state $|\{n\},\{{\hat n}\}\rangle$, with $\ncol=\{\nhat +\delta_K-\delta_L\}$,  where $\delta_K$ is unity for mode $K$ and zero otherwise.    Working in the near-horizon region, where the interactions should be strongest and the kinematics is that of Minkowski space, the tree-level approximation to this amplitude takes the form
\eqn\gravamp{\calA(p,|0\rangle_{in}; p',\ncol,\ncolh) = {i\over M_p^2} \int {d^4 q\over (2\pi)^4} \langle\ncol,\ncolh|\Delta T^{\mu\nu} (q) |0\rangle_{in} {1\over q^2} \langle p' |{\bar T}_{\mu\nu}(-q) |p\rangle}
where we use the bar notation
\eqn\bardef{{\bar T}_{\mu\nu} = T_{\mu\nu} - \hf g_{\mu\nu} T\ }
and $q$ is the momentum transfer.

At first sight it appears that these interactions are enormous, as a result of the large blueshifts -- the third term of \fdstexp\ makes a huge contribution like in \fdT.  However, one must also include the contribution of the modes {\it inside} the horizon.  Thus, for the combined contributions, we have an expression of the form
\eqn\Tmat{\langle\ncol,\ncolh| :T_{\mu\nu} (x): |0\rangle_{in} = \langle\ncol,\ncolh| \left[t_{\mu\nu}(v^*_K,v_L) b^\dagger_K b_L + t_{\mu\nu}(\vhat^*_L,\vhat_K) \bhat^\dagger_L \bhat_K\right] |0\rangle_{in}\ .}
For a state of the form \prodH, this becomes 
\eqn\Teval{\langle\ncol,\ncolh| :T_{\mu\nu} (x): |0\rangle_{in} = C{\sqrt{n_K(n_L+1)}\over \gamma_L}
e^{-\beta_HE(\ncol)} \left[\gamma_Lt_{\mu\nu}(v^*_K,v_L) + {\gamma_K} t_{\mu\nu}(\vhat^*_L,\vhat_K)\right]\ ,}
where $\gamma_K$ is defined in terms of the frequency of mode $K$ by \gammafd.
Using \vfd, the expression in brackets combines to give
\eqn\tcancel{\left[\gamma_Lt_{\mu\nu}(v^*_K,v_L) + {\gamma_K} t_{\mu\nu}(\vhat^*_L,\vhat_K)\right]={\sqrt{1-\gamma_K^2}\sqrt{1-\gamma_L^2}\over \gamma_K} t_{\mu\nu}(v^{1*}_K, v^{2*}_L)\ .}
Thus this expression is pure negative frequency in Kruskal time, the appropriate time for the infalling observer.  When one evaluates \gravamp\ in this infalling frame, momentum conservation then implies that no large momenta enter the expression.

\subsubsec{Evolution on slices}

From the preceding argument one finds that the effect of the large stress tensor of the ultra-high energy outside modes is cancelled by that of the inside modes.  Indeed, this result could have been anticipated from another perspective, that of  ``nice slice" evolution, and this indicates a full reconciliation of the two viewpoints, with no evidence of a physical role for the ultraplanckian precursors of Hawking radiation.

\Ifig{\Fig\Krusk}{The Kruskal diagram for the Schwarzschild geometry, together with two kinds of time slices.  An explicit construction of the first kind (solid line) is given by \LPSTU: it consists of a segment of the hyperbola of constant $r=r_c$ for $U>V$, attached to the matching horizontal slice in the region $U<V$.  The second kind (dashed) consists of a horizontal segment for $U<V$, which terminates at the singularity.  Action of the Schwarzschild time translation $t\rightarrow t+\alpha$ on either of these slices generates a family of slices foliating the region outside the black hole, as well as the region at weak curvature inside the black hole.}{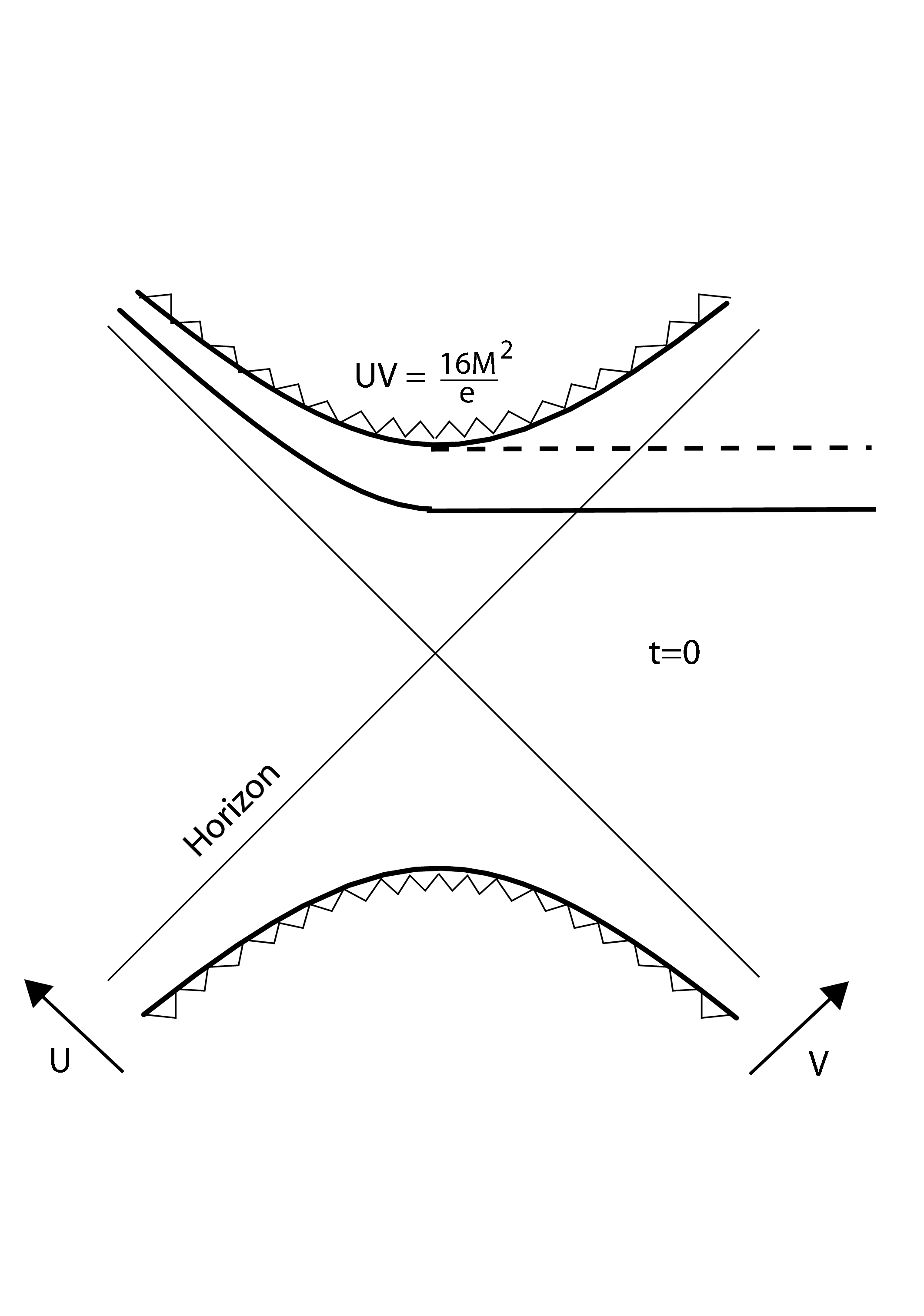}{4.5}

This latter perspective follows from constructing a family of spatial slices that is ``as smooth as possible," or ``nice," and describing evolution on these slices.  The Kruskal diagram for the Schwarzschild geometry is shown in \Krusk, which also shows two possible constructions of one slice in such a family.  The first type of slice avoids the planckian region near the singularity completely.  Or, with natural evolution tracking infalling observers, one would get a slice like the second, which enters this planckian region.  The intrinsic spatial geometry of the two slices is shown in Fig.~2 and Fig.~3.  Either of these slices agrees with the constant Schwarzschild $t$ slice as $r\rightarrow\infty$.  The full family of slices can be constructed by applying Schwarzschild time translations to either of the slices shown, {\it i.e.} acting by $t\rightarrow t+\alpha$.  Note that in Kruskal coordinates \kruskdef, these translations have action
\eqn\kruskxm{U\rightarrow e^{-\alpha/4M}U\ ,\ V\rightarrow e^{\alpha/4M}V\ .}

\Ifig{\Fig\nosing}{The intrinsic geometry of a member of the family of slices of \Krusk\ that avoids the strong curvature region}{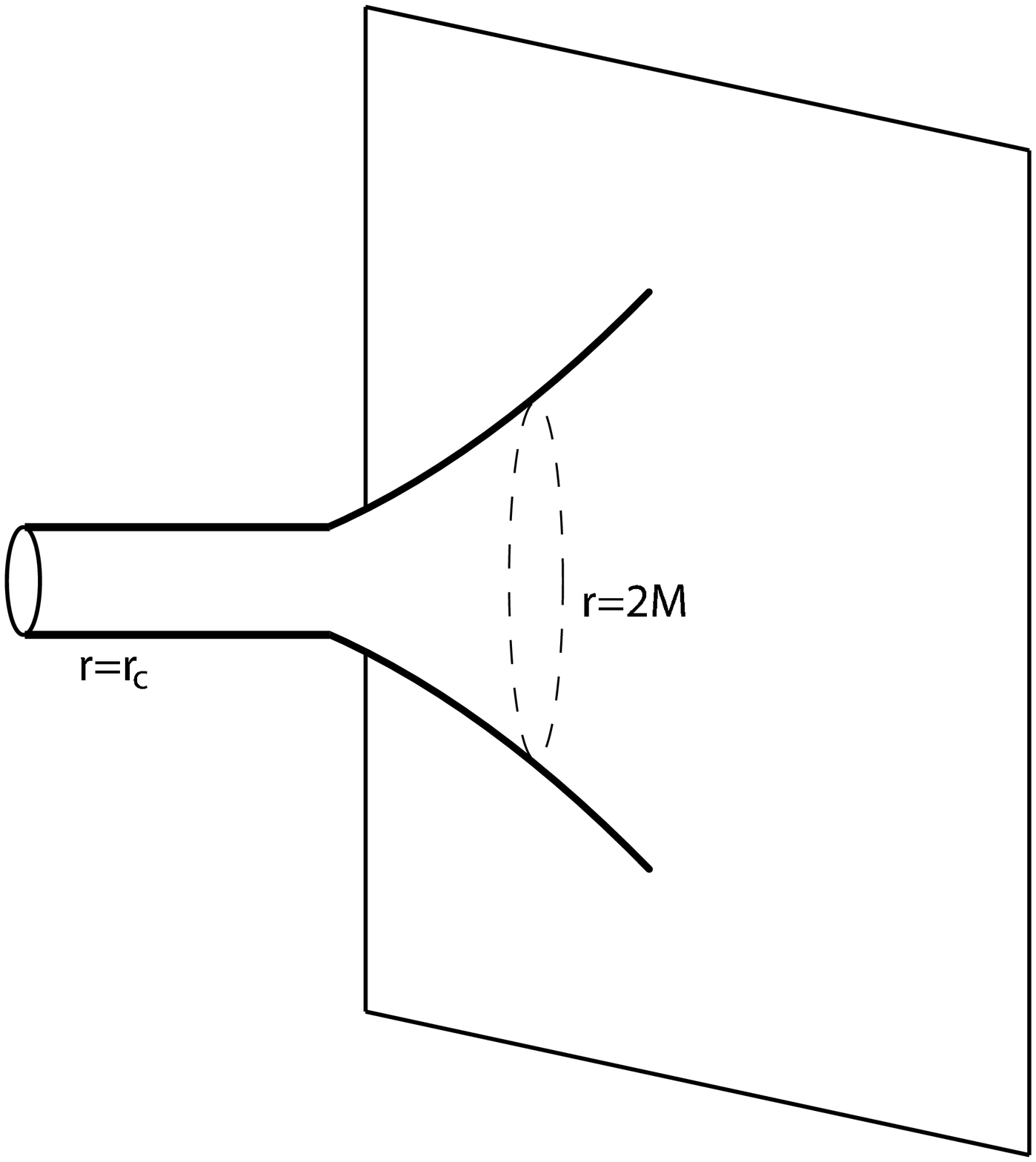}{3.3}

\Ifig{\Fig\singslice}{The intrinsic geometry of a member of the family of slices of \Krusk\ that intersects the singularity.}{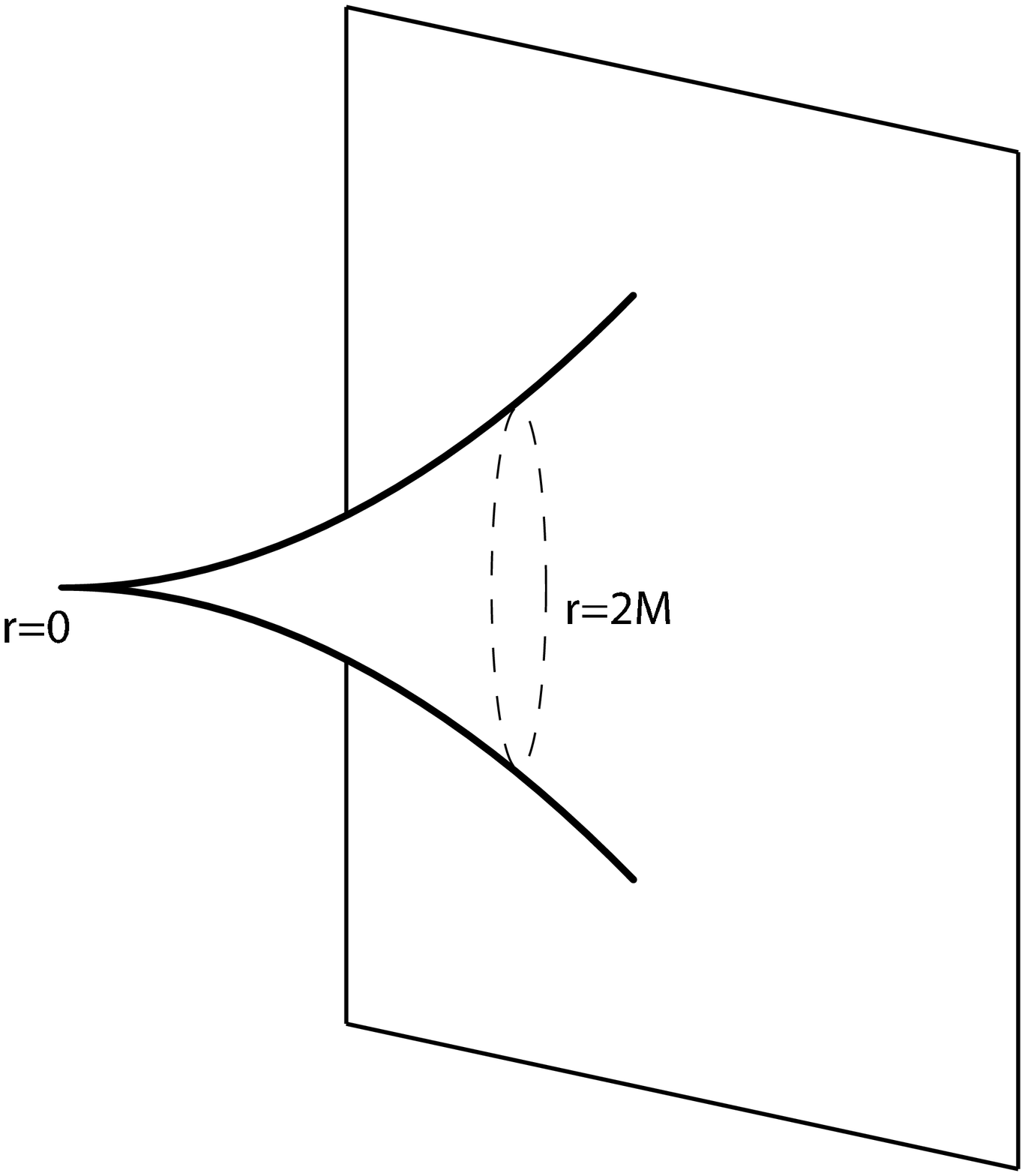}{3.3}

With either construction, it appears that evolution is smooth and causal in the vicinity of the horizon.  Moreover, 
a defining property of the Hawking state can be taken to be that it is annihilated by operators corresponding to positive frequency wavepackets that have support near or outside the horizon and that have frequencies $\gg 1/M$.  This follows from adiabaticity, in an extension of the argument given by \LPSTU\ -- if the initial state is vacuum, evolution on our slices in the vicinity of the horizon and outside is adiabatic for energies $E\gg 1/M$.  While this adiabaticity would break down near the singularity, locality and causality would seem to predict that no consequence of this could be seen outside the horizon.  

Such an argument thus indicates that the infalling observer sees a vacuum for the high energy modes.  One might have expected this to clash with the description appropriate to the outside observer, but the derivation of the effects of ultraplanckian modes of the preceding subsection shows that this is not the case, due to the cancellation between the effects of such modes inside and outside the horizon.  This clearly indicates that from a smooth-slice perspective, an essential dependence of Hawking radiation on ultraplanckian effects\refs{\tHooftRE,\tHooftFN,\tHooftFR,\tHooftUN\VerlindeSG-\SchoutensST} is a fake; on any given slice the state can be defined by the condition that it be the vacuum with respect to high-energy modes as seen by an infalling observer.  

\subsec{Wavepacket evolution}

While consistent and apparently correct, the above discussion leaves some puzzles.  First off, it seems odd, for example from the perspective of locality, that the effect of an ultraplanckian wavepacket inside the horizon could cancel the effect of one outside the horizon.  Moreover, the above arguments apparently further cement the case for information loss, and consequent paradox.  One can better understand these issues by examining detailed evolution of  wavepackets comprising the Hawking radiation.

Consider a quantum of Hawking radiation that leaves the black hole at some retarded time $u_0$.  The typical quantum will have 
energy $\sim T_H$ at infinity, and thus the corresponding wavepacket extends over a retarded time interval $\Delta u\sim 1/T_H\sim M$.  As this wavepacket is traced back to its origin near the horizon, its frequency blueshifts and its width correspondingly contracts, as seen in the infalling observer frames.  The essential characteristics can be understood by ray tracing.  Considering for example the evolution of the ray $u=u_0$ on the slices described in the preceding section, we see from the definition of Kruskal coordinates, \kruskdef, that the distance of the ray to the horizon decreases as 
\eqn\kruskray{U\sim Me^{-(u_0-t)/4M}}
on slices at early times $t$.  A key point is that in the near-horizon region, $U\roughly<M$, the width of the wavepacket is comparable to its distance from the horizon:
\eqn\wpwidth{\Delta U\sim e^{-(u_0-t)/4M} \Delta u\sim  Me^{-(u_0-t)/4M}\ .}
The inside wavepackets have a similar behavior.  Thus for $u_0-t\gg M$ the inside and outside wavepackets have neither cleanly separated from the horizon nor from one another.  This is one way to understand how their influence can cancel.  Only on slices with $t\roughly>u_0$ do the inside and outside wavepackets cleanly separate from one another and from the horizon. At this point the outside wavepacket will have an energy comparable to its asymptotic value, $\sim 1/M$.  Correspondingly, the inside wavepacket will descend into the strong curvature region.

In summary, this description reinforces the picture of the preceding subsection.  The infalling observer sees modes with frequencies $\gg 1/M$ in their vacuum; if one attempts to describe corresponding observations in terms of the ultraplanckian modes, the inside and outside ultraplanckian wavepackets overlap and the na\"\i vely large ultraplanckian effects cancel.  Conversely, as described on our smooth slices, only modes with frequencies $\roughly<1/M$ are appreciably excited.  This happens when the inside and outside modes separate, which happens when their frequencies reach $\calo(1/M)$.  Thus interactions only occur with Hawking modes once they become real outgoing particles, typically of energies $\omega\sim 1/M$.

\newsec{Nonlocality and information retrieval}

The preceding discussion appears to only tighten the paradox; apparently within the current framework of quantum field theory and general relativity there is no consistent description of black hole evolution.  In particular, the hypothesized relevance of ultra-high energies\refs{\tHooftRE,\tHooftFN,\tHooftFR,\tHooftUN\VerlindeSG-\SchoutensST} in the derivation of Hawking radiation appears to have been   ruled out.  This suggests the need for some new physical ingredient.

\subsec{Nonlocality in gravitational dynamics}

The step from classical physics to quantum physics similarly arose from crises, and required assumptions outside of the bounds of classical physics.  Likewise, here one might expect that a new assumptions are needed for a consistent description of black holes, that are apparently not derivable from the framework of quantum field theory and general relativity, and possibly don't even follow directly from string theory.

Beginning with the work of 't Hooft, that of ref.~\BHMR, and  Susskind and others' work on the holographic principle, it has seemed likely that the critical new assumption involves some form of nonlocality.  However, several things are needed to extend these ideas.  These include statements both of the mechanisms for such nonlocality, and of the domains in which such nonlocality are relevant.  Moreover, one needs to understand how such nonlocality could resolve the paradox.

It has long been suspected that the extended nature of strings gives a mechanism for nonlocality.  However, refs.~\refs{\GiLione,\GiLitwo,\SGloc} suggested an alternate mechanism for nonlocality, namely that it is simply {\it intrinsic} to the description of strong gravity.  Specifically, locality can be phrased in quantum field theory by the statement that if we divide a spacelike slice into two non-overlapping regions, the Hilbert space describing field configurations decomposes into a corresponding tensor product, or equivalently by the statement that observables commute at spacelike separations.  However, it seems that there must be something wrong with these statements in sufficiently extreme situations.  For example, quantum field theory measurements create or annihilate particles; if one tries to measure two particles of sufficiently high total energy in a small enough region, the backreaction of the created particles strongly deforms the metric and causal structure of the region.  In such circumstances there is no clear statement of locality, and a very reasonable proposal\refs{\GiLione,\GiLitwo,\SGloc} is that there is simply no local description of such a situation.  

Indeed, one can investigate the relative roles of gravitational and string nonlocalities through study of high-energy scattering\refs{\SGloc}.  In this context there is no evidence for intrinsically stringy nonlocalities associated with creation of very long strings, but one does find indications for modifications to a local description due to gravitational effects,  in particular at the gravitational radius given by the center of mass energy.\foot{There remains a possibility of intermediate effects, arising from macroscopic tidal string excitations\SGloc.}

These considerations suggest a general {\it nonlocality principle}, stating that in such circumstances where gravity is strongly interacting, a complete description of the fundamental degrees of freedom is not local over the strong gravity region.  A reasonable assumption is, however, that their description is still quantum mechanical, and in particular results in a unitarity S-matrix in the scattering context.  A nonlocality principle stating that in certain circumstances one cannot describe physics in terms of local degrees of freedom bears a strong analogy to the uncertainty principle, which states that in certain circumstances one cannot describe physics in terms of classical phase space degrees of freedom and instead must resort to the quantum-mechanical wavefunction.  Thus, the dynamics underlying such a nonlocality principle could be anticipated to be a substantial departure from quantum field theory and general relativity, and the inference of the relevant mathematical and physical framework would be analogous to the invention of quantum mechanics.

In the absence of such a framework, one must by necessity be somewhat heuristic and 
can at best give a rough parametrization of the regime in which locality should fail.  Statements of locality bounds, given in \refs{\GiLione,\GiLitwo,\SGloc} provide one criterion to enter the domain where locality is hypothesized to fail.  Specifically, in the context where one attempts to study a pair of quanta, described by wavepackets with approximately definite positions $x$ and $y$ and momenta $p$ and $q$, a rough criterion for validity of $D$-dimensional locality is
\eqn\locbd{|x-y|^{D-3}\roughly>G_D |p+q|\ ,}
and violation of this bound is proposed to indicate its breakdown.
Similar statements can be given in the multi-quanta case\SGloc.

An important test of such a nonlocality hypotheses is that it offers the possibility of escape from the black hole paradox.  Certainly nonlocal physics holds good potential, since locality prevented the escape of information from the black hole.  But once one attempts to write criteria for such nonlocality to be operational, it will only be relevant to the black hole problem if such criteria apply to the black hole dynamics.  Given the above statements, particularly \locbd, we can seek to identify a situation which produces a large energy invariant, and thus a possible rationale for locality, and the arguments that produced the paradox, to fail.  It is also possible that strong gravity produces other mechanisms, {\it e.g.} similar to envisioned in \BHMR, which may not be directly parametrized by \locbd.

Before turning to the black hole, it is important to make a basic assumption in the discussion explicit:  
Lorentz invariance is taken to be exact, to arbitrarily high boosts.  This stands in contrast to work suggesting modified dispersion relations, maximum velocities, or preferred frames\refs{\MagueijoAM\Amel\Amelino\Kowal-\HossenfelderCW}, as well as approaches based on an explicit cutoff\refs{\Jacob}.  However, it appears both plausible and for example in accord with our knowledge of string theory.  In particular, it seems reasonable that we can describe a particle with ultraplanckian momentum by viewing a particle at rest from a sufficiently boosted frame.  The semiclassical approximation to the geometry of such a particle should be just the Aichelburg-Sexl metric\refs{\AiSe}.
While we assume that there is nothing wrong with the {\it kinematics} of such a description, as we've argued, we do expect that ultraplanckian boosts lead to important effects on the 
 {\it dynamics}. In particular, the cross-section for a collision of this particle with another, say at rest, will grow with the energy due to black hole formation\refs{\Penrose\EaGi-\GiddingsXY}.   This is viewed as a special case of the nonlocal unitary dynamics of strong gravitational physics.

\subsec{Nonlocality in black hole evolution}

To escape the paradox,  a rationale for the relevance of new nonlocal physics, and breakdown of the semiclassical approximation, should  be identified.  In particular, one might begin by trying to identify a large energy invariant pertinent to the dynamics of the Hawking radiation.

Section three argued that interactions of the Hawking radiation are not present until the Hawking modes attain separation $\sim M$ from the horizon, and thus have energies (measured for example by local observers) of order $1/M$.   In a sense the emitted particles are only created at this time and don't have an independent existence before this.
These facts apparently undermine  motivation for the transplanckian frequencies of Hawking's original derivation to justify new nonlocal effects.

\subsubsec{Invariants and estimates}

However, even taking this into account, one can  identify other large invariants.  Consider a particle with asymptotic energy $E_1$ that falls into the black hole at time $t=t_1=0$.  We are interested in its possible influence on late-time Hawking radiation.    A given Hawking mode can begin to have interactions at the time it begins to leave the region of the horizon; consider a mode that does so at time $t_2$.  Let us first describe the process in terms of the time slicing that avoids the singular region.

For large $t_2$, the $t=t_2$ slice is highly boosted relative to the $t=0$ slice shown in \Krusk.  However, this can be undone;
a description of the Hawking mode as seen by an infalling observer in the vicinity of its creation can be found by acting with the translation $t\rightarrow t-t_2$ which brings the time slice labeled by $t_2$ to the one pictured in \Krusk.  Through \kruskxm, one finds that this boosts the energy of the infalling particle to 
\eqn\booste{E_1'\sim E_1 e^{t_2/4M}}
with respect to the frame of the observer at $t=t_2$.  Since the outgoing Hawking mode has energy $E_2\sim 1/M$ in this frame, there is a large energy invariant, 
\eqn\largeinvt{s \approx E_1E_2 e^{(t_2-t_1)/4M}\ .}
This is independent of the particular time slicing, and results directly from comparison of the energies of the particles in the near-horizon Rindler region.

Next consider the relative separation of the modes.  If the metric of \Krusk\ were flat, this would also be exponential in $t_2-t_1$ and the locality bound \locbd\ would not be violated.  In the black hole metric there are different ways to compute this separation.  One measure is the distance along the spatial slices.  This is trivially estimated: if the hyperbolic part of the slice lies at $r=r_c$, the distance is\foot{For the remainder of the paper the focus is on $D=4$, though extension to larger $D$ is straightforward.} 
\eqn\distance{\sigma \approx \sqrt{2M\over r_c} (t_2-t_1)\ .}
Indeed, the statement that the separation at most grows as $t_2-t_1$ is slice independent.
On the second type of slice, shown in \singslice, the separation is possibly even less, $\calo(M)$, but there one needs planckian physics to describe the states.  

\Ifig{\Fig\intmodes}{A representation of the situation described in the text, on a slice avoiding the singularity.  Both a Hawking mode and its inside counterpart are represented, in the process of separation from the horizon.  A mode that entered the black hole at an earlier time is boosted relative to these modes by an amount exponential in this time difference, whereas its distance along the slice is linear in the time.}{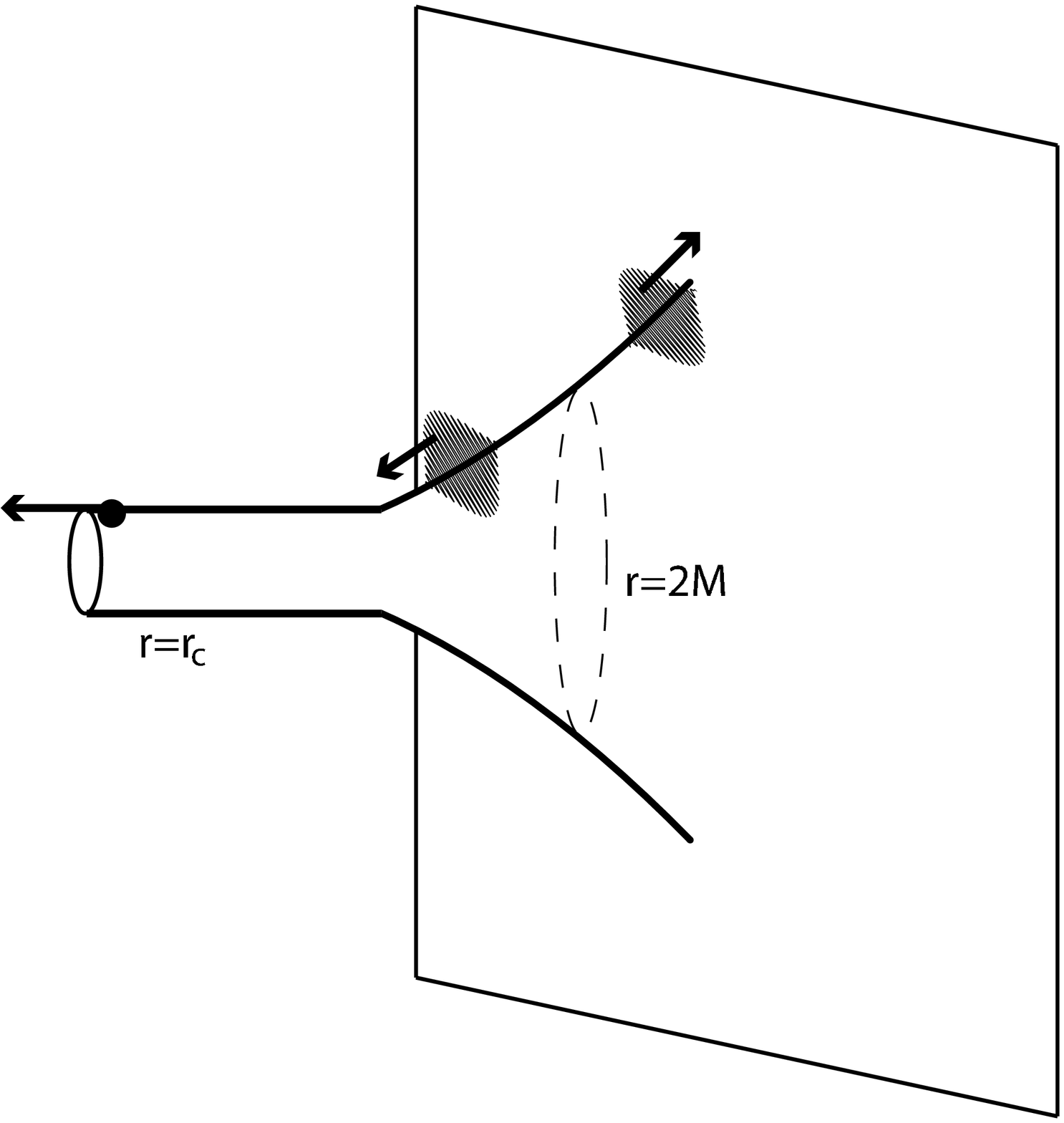}{4}

An essential question is whether one can think of the Hilbert space on the the time slice $t=t_2$ as decomposing into separate factors describing the infalling mode and the outgoing Hawking mode; see \intmodes.  As we've described, this is what is required to give a clear argument that one can trace over states of infalling particles to find a mixed outside density matrix.  
Comparison of the invariants \largeinvt\ and \distance, in accord with \locbd, suggests that such a decomposition breaks down.  However, since the background is not flat, one might question applicability of the bound \locbd.  This can be assessed by investigating the structure of gravitational amplitudes.

\subsubsec{Amplitude analysis}

The preceding rough estimates can be fleshed out, in an essentially slice-independent fashion, by investigating the structure of the amplitudes governing the modes' interaction.  Consider the tree-level amplitude for graviton exchange between the infalling quantum and the outgoing Hawking particle, which has a structure analogous to \gravamp.  Specifically, using the flat space kinematics, one finds an amplitude of magnitude
\eqn\gravampn{ {\cal A}_{tree} \sim {i\over M_p^2}\int d^4x_1 \sqrt{-g} \int d^4 x_2\sqrt{-g} T_{1\mu\nu}(x_1){\bar T}_2^{\mu\nu}(x_2) G(x_1,x_2)}
where $T_1$, $T_2$ are the stress tensors of the infalling and Hawking particles, respectively, and $G(x_1,x_2)$ is the massless scalar Green function.


The large center-of-mass energy squared, \largeinvt, enters this expression through the product of stress tensors, as easily follows from the flat-space kinematics of the Rindler region.  Consider analyzing the problem in the center-of-mass frame, which is reached through a boost, {\it i.e.} a $t$ translation.  In this frame both particles have exponentially large and oppositely directed momenta. 
To ascertain the possibility of a large amplitude, the Green function needs to be estimated.

If the geometry were simply flat space, this Green function would take the form 
\eqn\flatgreen{G(x_1,x_2)\propto{1\over  (x_1-x_2)^{2}}\ . }
In the center of mass frame, the flat space separation between $x_1$ and $x_2$ at the time corresponding to creation of the Hawking quantum 
would be exponentially large in $t_2-t_1$, and would not produce a large amplitude.  However, $G$ needs to be computed in actual black-hole geometry.  The first-quantized representation of the massive propagator is
\eqn\mgreen{G(x_1,x_2) = \int_0^\infty {dT} \int_{x_1}^{x_2} \cald X \exp\left\{{i\over 2} \int_0^T d\tau ({\dot X}^2 - m^2)\right\}\ ,}
in which we can take the limit $m=0$ for the present case.   In a saddlepoint approximation, \mgreen\ can be computed in terms of the geodesic(s) from $x_1$ to $x_2$.  For example, in flat space, spacelike $x_1-x_2$ yields \flatgreen.  

The problem of finding such geodesics was discussed for the similar problem of AdS-Schwarzschild in \refs{\FHKS}.  Consider first the case where the modes are traveling along the same ray from $r=0$, so that the separation  between $x_1$ and $x_2$ is purely radial.   In this case, as in \FHKS, we find there is no radial geodesic in Schwarzschild that connects the two points.  In effect, the extremal trajectory gets dragged into the singularity.  However, if some planckian physics removes the singular behavior, one would expect an extremal trajectory to exist.  As a simple model, suppose that the singularity is replaced by a simple boundary condition that cuts off the geometry at a critical radius $r=r_c$.  In that case, the extremal curve (with constraint $r\geq r_c$) will hug $r=r_c$ for much of the span between $x_1$ and $x_2$.  Thus its length takes the form \distance.  Even in pure Schwarzschild, as pointed out in \FHKS, there will be spacelike geodesics with non-zero angular momenta connecting typical spacelike-separated points, also with lengths $\propto t_2-t_1$.  Thus, while all approximations are not under control, we motivate an expression of the form
\eqn\masslessg{G(x_1,x_2)\sim {1\over \sigma^2}}
with $\sigma$ given in \distance.\foot{Moreover,  the picture \intmodes\ suggests that gravitational field lines would be squeezed by the geometry of the slice, perhaps producing even stronger gravitational effects at a given distance.}

Of course, a complete calculation of the Green function, and tree level amplitude, requires knowledge of the planckian regime near the singularity.  But this rough estimate indicates the structure such a calculation could take.  Moreover, the estimated tree-level amplitude is large in the regime resulting from the more na\"\i ve guess \locbd.  Growth of the amplitude is indicative that the tree-level approximation is not complete; when it becomes sufficiently large one must include other diagrams to compute the full amplitude.  This brings us into a domain where the proposal plausibly applies,  that  amplitudes are unitarized in an intrinsically nonlocal way when gravity becomes strong.  

Thus, while a complete calculation can't be given, this suggests  what is needed.  Specifically, the gravitational interaction between the infalling mode and the outgoing Hawking particle is potentially large, but depends on strong gravitational physics.  This should be contrasted with the opposite alternative:  the interaction can be calculated in a controlled approximation, and is small.  The latter situation would lead to failure of quantum mechanics and paradox.  But the former, in accord with the conjectured nonlocality, indicates how the infalling particle and the outgoing Hawking mode could avoid being independent degrees of freedom, and this suggests a way that quantum mechanics can be preserved and paradox avoided.

This argument is quite general.  In particular, it is independent of specific constructions of slices and relies only on the kinematics of Schwarzschild geometry.  Any attempt to describe both the infalling matter and outgoing Hawking modes simultaneously on a single spatial slice, without involving planckian physics, as is necessary for the general form of Hawking's argument given in section two, encounters very large relative boosts which we have argued can activate proposed nonlocality in strong gravitational physics.  

The detailed estimates above involve one possible symptom of the requisite nonlocality; there may be other descriptions or even other sources.  For example, one could ask about a description using the ``singular slices," shown in \singslice. However, here even a  description of the information that has fallen into the black hole would require full understanding of planckian dynamics.  In such a description it is both less clear how to describe the information, and what is the large invariant that could parameterize the nonlocality.   One possibility is that it is  be related to $M/r_2^{D-3}$, where $r_2\sim M$ is the radius at which the Hawking particle is produced.  But a planckian resolution of the singularity would have other strange features.  For example, if information is preserved and the geometry indeed terminates in the strong curvature region, as suggested in \singslice, then in some sense information would be described as propogating in a superluminal fashion.  (For related discussion see \BHMR.)  It may be that whatever physics resolves the singularity has intrinsically nonlocal features and/or predicts new kinds of instability that play an important role.

One can also ask for which modes the proposed nonlocality would be operational.  If one for example takes the relation \locbd\ at face value, it indicates that the failure of a decomposition into independent Hilbert spaces would occur for\foot{It is conceivable that some string effects could lead to nonlocality  at even lower blueshifts, thus shorter times.  But, following the discussion of \SGloc, the gravitational effects are likely to retain relevance.} 
\eqn\locbdt{t_2-t_1\roughly> 4M\log \left[{1\over E_1 E_2} \left({2M\over r_c}\right)^2 (t_2-t_1)^2\right]\ ,}
or, with $E_1\sim E_2\sim 1/M$ and $r_c\sim M$, 
\eqn\locbte{t_2-t_1\sim 16M \log M\ .}

The reasoning of this section shares some common elements with the discussion of \LPSTU.  However, there are important differences in the kinematical setup, and moreover in the essential mechanism for nonlocality.  Ref.~\LPSTU\ argued that this should arise from a long string stretched between the infalling matter and the outside region, but moreover were uncertain about having found a gauge invariant effect\Polchrev.  It is hard to understand how such long stretched strings could contribute to important nonlocality in this context, but apparently not contribute important nonlocal effects in the context of high-energy scattering\SGloc.  (One must make the plausible assumption that the same mechanism of nonlocality is important in these different backgrounds.)

In summary, it doesn't appear possible to make Hawking's arguments for information loss without 
reference to very large relative boosts and/or planckian physics. Thus, while our description of the ``singularity states" is limited, the description we have motivates the entrance of strong gravity and nonlocal phenomena.  There may be other different descriptions of such  phenomena, in different pictures or frames.
While even opening a window for resolution of the paradox appears to be progress, given the absence of other viable alternatives, clearly a deeper understanding would be desirable.

\subsec{Discussion: holography, complementarity, retention time}

In short, there does not appear to be a convincing case that the Hilbert space of states on a spacelike slice factorizes into separate tensor factors in sufficiently extreme circumstances, in particular described by the locality bound \locbd.  If there is no reason to trust local quantum field theory in this situation, there is apparently not  good reason to argue that the modes inside and outside the black hole have independent existence.  Put more precisely, for such modes a decomposition of the form \hilbprod\ hasn't been justified, and according to the nonlocality principle, is proposed not to exist.
Such an argument suggests that it doesn't make sense to think of  information ``lost" to the interior of the black hole.
This hypothesis is just what is needed as a loophole in Hawking's argument, and a resolution of the paradox.  This breakdown of a local description is proposed to be a fundamental feature of strong gravitational physics, which in essence provides the mechanism for nonlocality.

Thus, such nonlocal gravitational physics could serve as a deeper explanation of holography.  If despite the nonlocal dynamics, a valid semiclassical description of observations of the infalling observer still exists, this 
 picture could also provide part of the explanation of black hole complementarity \refs{\tHooftRE\tHooftFR\tHooftRB-\SusskindIF}.   These earlier references argued that
there is no way to compare information of infalling observers to that in the Hawking radiation, and thus that they may be represented by complementary descriptions.   Nonlocal gravitational physics along the lines parametrized by the locality bound may give a deeper rationale for the failure of their observations to commute, analogous to noncommutativity of position and momentum in quantum mechanics.  
 
If we consider a bit of information carried into the black hole by an infalling quantum, an important question is when its information could first be present in the Hawking radiation, or alternatively, which states in the outside Hilbert space do not have an independent description that is compatible with the inside description of this bit.  At present we lack the tools to fully answer this question; one needs a more complete understanding of the principles of the underlying nonlocal physics.  But \locbdt\ and \locbte\ are suggestive as the relevant time scale for the delay in accessibility of the information.   The time scale \locbte\ has been previously identified as important, based on apparently different logic.

Another open question for future work is to try to infer how precisely the information would be parameterized in the outgoing density matrix that replaces \densmat. Motivated by our earlier discussion, this could arise from non-zero off-diagonal elements in the density matrix, as in \extwo, as well as possibly modified diagonal components.  Actual computation of these elements appears to require a complete description of strong quantum gravity, but there may be means to make estimates.

Lastly, another proposal for for the fate of information in black holes is the ``final state" proposal of Horowitz and Maldacena\refs{\HoMa}, which suggests that the state at the singularity is unique, and thus devoid of information.  But in order to explain such behavior, some nonlocal physics would be required to relay the information outside the black hole.  The present discussion of nonlocality is a proposal of such a mechanism. Specifically, present arguments suggest that the inside and outside Hilbert spaces don't have independent existence in certain contexts.  For example, the  form of complementarity described above, and the idea of the locality bound, suggests that if there are complementary pictures, in an ``outside" picture, an independent inside Hilbert space loses meaning.  Despite the proposed failure of independence of inside and outside Hilbert spaces, one might nonetheless attempt a mathematical description that describes both simultaneously, with the understanding that such a description would be redundant.   One possible way to understand this redundancy could be through
application of a final internal state boundary condition as in \HoMa.
Conversely, if there is a valid ``inside" complementary picture, dynamics seen by an infalling observer may look different from that enforced by such a boundary condition.

\newsec{Conclusion}

The sharpness and fundamental character  of the black hole information paradox suggests that it should help guide our progress to a more complete quantum mechanical and gravitational physics.  A sense that the resolution lies in some nonlocality in physics has grown, and some suggested characterizations of this proposed nonlocality have been given, in particular through the ideas of black hole complementarity and holography.  However, lacking in these statements has been a description of the mechanism of such a basic nonlocality, and a description of when this mechanism is operable.  In short, without identifying a  loophole in Hawking's derivation of information loss, the paradox remains.

Considerable effort has focussed on the possibility that the kinematics of the ultra-planckian modes in Hawking's derivation could play an important role, and indeed might lead to a dynamical explanation of how infalling information is imprinted on outgoing Hawking radiation.  This paper has examined such a scenario more closely, from the point of view of the dynamics on smooth slices\refs{\LPSTU,\Jacob}.  It has in particular described aspects of how the ``nice slice" expectation that there is no relevant ultra-planckian dynamics at the horizon can be reconciled with the apparent relevance of such dynamics from the viewpoint of Hawking's derivation.  Specifically, it appears that the outgoing Hawking modes do not have non-trivial interactions, and in fact don't attain an ``independent existence," until they reach energies (as seen by a nearby freely falling observer) $E\sim 1/M$, and begin to leave the horizon region.  Before that the state describing these modes and their counterparts inside the horizon looks just like the vacuum to an infalling observer.  These arguments appear to weaken  suggestions that ultraplanckian dynamics of Hawking modes alone is responsible for the breakdown of Hawking's derivation of information loss, and consequently tighten the paradox.

If nonlocality is indeed the resolution of the paradox, this paper has instead argued that in this picture the nonlocality should allow infallen information to influence the Hawking modes around the time they become real and separate from the horizon.  A suggested rationale for such nonlocality has been given:  if one attempts to draw  a spacelike slice that intersects both the  infalling modes and the outgoing Hawking modes, to compare their information, one finds a relative boost between these modes that grows exponentially in the time between infall and emission.  This, together with the fact that these modes are confined to the black hole region, which na\"\i vely has size $\sim M$ (though may have greater length scales in its internal geometry, as discussed), suggests that an attempt to simultaneously describe both sets of modes encounters strong gravitational dynamics.  It has been argued\refs{\GiLione,\GiLitwo,\GMH,\SGloc} that such dynamics is inherently nonlocal; there may be other realizations of such nonlocality.

This discussion doesn't refer directly to properties of string theory.  One might have expected that nonlocality alternately arises from the extended nature of strings.  However, ref.~\refs{\SGloc} has argued that investigations of high-energy scattering in string theory show no evidence of such nonlocality on the na\"\i ve distance scales $\propto l_{st}^2 E$.  While there could be a possible role for other nonlocalities at shorter distances, due to tidal excitation dynamics (and they could in fact fit into an analogous discussion of information in Hawking radiation), it appears that an ultimate limit on locality comes from strong gravitational dynamics.  Moreover, this genericity mirrors that of the paradox.

In some respects it seems dissatisfying to push the question of the fate of information into a domain where no present calculation can answer it conclusively.  However, in the final analysis this is precisely what is needed.  The paradox arose from Hawking's claim that information destruction could be derived in a controlled approximation from known physics.  However, we have described how
large invariants in the kinematics of infalling and Hawking modes suggest a loophole through which nonlocal dynamics could be operative, and thus indicate a possible way to avoid the claim that led to the paradox.
The lack of other viable alternatives for its resolution adds weight to any such alternative, no matter how counterintuitive.  

The reader who is disturbed by the lack of a precise mathematical description of principles of nonlocal physics, and of their applicability to the black hole information paradox, should consider the analogy of the transition from classical physics to quantum mechanics.  Faced with the crisis of the classical instability of atomic matter, at first the best Bohr could do is invent rather ad-hoc rules within a classical framework: in his words, he proposed resolving the paradoxes of atoms by means of a ``hypothesis for which there will be given no attempt at a mechanical foundation (as it seems hopeless) ...\Bohr."  Moreover, ``This seems to be nothing else than what was to be expected as it seems rigorously proved that the mechanics cannot explain the facts in problems dealing with single atoms\Bohr."
Acceptance of these hypotheses was motivated by their removing the crisis, without producing contradiction with known phenomena.\foot{Prediction of experimentally verifiable aspects of spectra was of course also important; alas, we cannot yet rely on a parallel.}  While arguments parametrizing the boundaries of classical physics exist, for example Heisenberg's microscope, and these capture aspects of the framework of quantum mechanics, a full understanding of basic issues even for the hydrogen atom required development of the mathematical and physical apparatus of quantum mechanics.  One could not expect to give a complete account of the physics in a classical
framework.  

The present situation has certain parallels to this history.  Our crisis is the black hole paradox.  Apparently the rules of quantum field theory plus general relativity must be modified, but only in regimes that don't conflict with known phenomena.  A proposal motivated by general considerations of what can be measured in the context of dynamical gravity is that locality breaks down in certain extreme circumstances.  This proposal appears to explain how to avoid the crisis.  But, if this approach is correct, a full understanding of such nonlocal dynamics of black holes must await development of a fundamental mathematical and physical framework.  A complete understanding of the physics presumably cannot be attained merely within the framework of quantum field theory together with general relativity.  A qualitative leap must be made.  But, outlines of the possible nonlocality do appear, and in particular suggest a drastic reduction in the number of degrees of freedom of the theory, like for example those discussed in the context of holographic proposals.

This viewpoint certainly draws strength from the sharpness of the paradox.  And if another possible solution is found, it could certainly weaken this case.  However, the fact that no generally accepted resolution has been found in thirty years of crisis raises the value of any proposal that offers a consistent logical alternative. 

\bigskip\bigskip\centerline{{\bf Acknowledgments}}\nobreak

I would like to thank D. Marolf for explaining why he didn't believe the arguments of \GiLitwo\ and for other discussions, J. Hartle for important discussions and support,  S. Hossenfelder for  important questions and conversations,  J. Polchinski and J. Preskill for valuable discussions, and G. Horowitz for comments on a draft.  Parts of this work were carried out while visiting Nordita and the Galileo Galilei Institute for Theoretical Physics, both of whose hospitality is gratefully acknowledged, along with the partial support of INFN.
This work  was supported in part by the Department of Energy under Contract DE-FG02-91ER40618.


\listrefs
\end